\documentclass[11pt]{article}
\usepackage[margin=1in]{geometry}
\usepackage{graphicx}
\usepackage{amsmath,amssymb}
\usepackage{latexsym}
\usepackage{framed,multirow}
\usepackage{xcolor}
\usepackage{url}
\usepackage{hyperref}
\usepackage{newtxtext,newtxmath}
\usepackage[numbers,sort&compress]{natbib}

\begin{document}

\begin{center}

{\LARGE \textbf{Patient-centered visualization of multistage cancer treatment trajectories}\par}

\vspace{1.5em}

Laura Lackner$^{1}$,
Marius Bill$^{2}$,
Martin Bornhaeuser$^{2}$,
Karolin Trautmann-Grill$^{2}$,
Helena Klara Jambor$^{1,2,*}$

\vspace{1em}

{\small
$^{1}$University of Applied Sciences of the Grisons, DAViS, Pulvermühlestrasse 57, Chur, 7000, Switzerland\par
$^{2}$Universitätsklinikum Carl Gustav Carus an der Technischen Universität Dresden, Germany\par
$^{*}$Corresponding author: helena.jambor@fhgr.ch\par
ORCID: 0000-0003-3397-1842
}
\end{center}
\vspace{1em}

\begin{abstract}
Effective communication of multistage cancer treatment trajectories 
remains a major challenge, particularly for patients with limited 
health literacy. We present a patient-centered visualization approach 
for representing complex, phase-based oncology treatments, 
integrating principles from information visualization, user experience (UX) design, 
and cognitive psychology. Using acute myeloid leukemia (AML) as a case study, 
we developed two timeline-based representations: a static, visually 
simplified trajectory emphasizing structure and hierarchy, and an 
interactive variant with layered information. We evaluated both 
approaches in a quantitative survey, measuring comprehension
of treatment sequences, perceived confidence, and information quality. 
Results show that the static visualization significantly improves 
understanding and clarity, highlighting the importance of visual 
hierarchy, consistent encoding, and reduced complexity when 
communicating temporal medical processes compared to the baseline. In contrast, 
additional interactivity did not improve performance and 
introduced navigational overhead, suggesting that interaction must 
be carefully aligned with cognitive demands. Our findings contribute
to visualization research by demonstrating how patient-centered design
can improve the interpretability of multistage treatment trajectories. 
We derive design implications for temporal medical visualizations, 
emphasizing simplicity, structural clarity, and accessibility to support 
informed decision-making in clinical contexts.
\end{abstract}
\clearpage
\twocolumn

\section{Introduction}
The quality of healthcare directly impacts patient outcomes and is defined not only by safety, effectiveness, and timeliness, but also by patient centeredness and equity \cite{wolfe_institute_2001}. Equity requires that patients understand their diagnosis, prognosis, and treatment options in order to participate in shared decision making \cite{odisho_patient-centered_2017,the_cochrane_collaboration_decision_2014}. In practice, however, this requirement is often not met. Patients frequently struggle to comprehend medical information, particularly when it involves uncertainty, risk, and extended temporal processes, as is the case in oncology. 
Low health literacy limits patients’ ability to engage with their care and interpret medical information \cite{galesic_using_2009,lipkus_numeric_2007}. These challenges are amplified among vulnerable populations, including older adults, non native speakers, and individuals with lower educational backgrounds \cite{garcia-retamero_communicating_2013,houts_role_2006}. At the same time, clinicians tend to overestimate patients’ level of understanding, further widening communication gaps \cite{houts_role_2006}. As a result, patients typically recall only a fraction of the information communicated during consultations, often no more than half and in many cases substantially less \cite{ley_memory_1979,jenkins_what_2011,kessels_patients_2003}. This has direct clinical consequences, as low satisfaction with health information is associated with increased anxiety and poorer outcomes \cite{goerling_information_2020}. These challenges are particularly pronounced in oncology, where treatment pathways are complex, multistage, and extend over long periods \cite{von_bonin_clonal_2021}. Patients must understand sequences of interventions, transitions between phases, and associated risks. However, clinical communication remains largely based on verbal explanations and text based materials, which are poorly suited to conveying temporally structured information. This can lead to reduced comprehension, lower confidence, increased decisional conflict, and potential non adherence to treatment \cite{husson_relation_2011,the_cochrane_collaboration_decision_2014}.

Visualizations offer a promising approach to address these challenges by transforming complex information into structured and interpretable representations \cite{spiegelhalter_visualizing_2011}. In particular, timeline and trajectory visualizations can make temporal relationships explicit, enabling users to better understand sequences, durations, and transitions. While such approaches are well established in clinical systems for expert users, their application to patient communication in oncology remains limited and underexplored. Moreover, the role of interaction in this context is not well understood, as it may either support or hinder comprehension depending on its design.

In this work, we investigate patient centered visualization of multistage cancer treatment trajectories. Using acute myeloid leukemia as a case study, we develop and evaluate two approaches: a visually optimized static timeline emphasizing clarity and structure, and an interactive variant with layered information. Through a quantitative study, we assess their effects on comprehension, confidence, and perceived information quality.
\section{Related work}
\subsection{Visualizing Temporal and Multistage Processes}
Representing temporal processes is a central challenge in information visualization, particularly when data involve sequences, durations, and hierarchical phase structures. Timeline visualizations provide a fundamental approach for encoding such data, enabling users to interpret temporal order, overlap, and duration \cite{aigner_visualization_2023}. However, as processes become more complex standard timeline representations can become difficult to interpret, especially for non-expert users. This is particularly relevant in domains such as healthcare where visualizations are critical for decision-making. Here, temporal processes are not only complex, and include point as well as interval events, but also merges different granularities of calendar time and relative time lines, both relevant for stakeholders \cite{cousins_visual_1991}. A different angle of timeline visualization focusses on the personal trajectories. LifeLines introduced a general framework for representing multi-faceted longitudinal health data with a single view, where multiple timelines, event icons, and visual encodings (e.g., color, thickness) enable users to identify patterns, anomalies, and relationships while supporting interactive filtering and rescaling for focused exploration \cite{plaisant_lifelines_1996}. This was further advanced in the TimeLine system that, for electronic health record data, enabled integration of clinical events, laboratory values, treatments, and reports into a multi-track temporal visualization designated to support interactive exploration for longitudinal patient representations \cite{bui_timeline_2007}. More recently, this concept was adopted in a visual analytics system for tumor board decision-making based on such longitudinal patient data. Specifically, this was developed for melanoma and supports comparison across patients and could improve the understanding of a patient history and decision-making in tumor boards \cite{pereira_visual_2023}. At present, most existing timeline and trajectory visualizations prioritize analytical depth for expert users rather than accessibility for patients. As a result, they often include dense encodings, multiple variables, or interactive exploration features that may hinder comprehension for non-expert audiences.
\subsection{Visualization for Non-Expert Users}
Visualization research is often oriented towards data exploration and advanced settings, however visualizations are also essential for non-expert audiences. Here, visualization must communicate clearly, especially in the health sector, to heterogeneous users, requiring designs that prioritize guided understanding, simplicity, and narrative over open-ended analysis and data completeness \cite{kosara_presentation-oriented_2016,chen_reflections_2020}. Building on this, narrative visualization has emerged as a promising approach to communicate complex, also medical information, to non-expert audiences by combining storytelling with interactive visual representations \cite{segel_narrative_2010}. In the medical setting, narrative visualization was for example implemented with a structured template for disease narratives, including one cancer type, and in evaluations improved engagement, understanding, and memorability compared to text-based formats \cite{meuschke_narrative_2022}. Enriching static formats with interactive elements, e.g., a slideshow or scrollytelling and a interactive navigation further shapes non-experts engagement into actionable change in user behaviour and confidence \cite{mittenentzwei_investigating_2023}. Importantly, this work demonstrated that design choices for user interactions directly affect usability and engagement. 

Another angle in medical communication is to integrate icons and pictograms to increase accessibility of non-expert users. For numerical medicall information, such as risks, replacing statistical charts with icon arrays and reducing the number of displayed options leads to markedly higher comprehension and notably increased satisfaction \cite{zikmund-fisher_communicating_2008}. This format works well in shared-decision making with patients, e.g. when communicating cancer survival under different treatment regimes. Another study also demonstrated that icons and icon arrays improve the accuracy of risk understanding, importantly that this was the case across both low- and high-numeracy groups, making them an effective tool for communicating medical risks to diverse patient populations \cite{galesic_using_2009}. Thus, for non-expert audiences narration, interaction and symbols all are strategies to improve comprehension of medical information and support informed decision-making. 
\subsection{Visual Aids for Patient Communication}
Visual aids are well established as effective tools for improving comprehension of complex information, particularly in the communication of risks, probabilities, and uncertainties \cite{spiegelhalter_visualizing_2011}. Visualizations like icon arrays have demonstrated benefits in health care communication, particularly for individuals with low numeracy or literacy \cite{galesic_using_2009,garcia-retamero_communicating_2013,odisho_patient-centered_2017}. As part of shared-decision aids, visual representations have been shown to improve patient knowledge, reduce decisional conflict, and support informed decision-making \cite{the_cochrane_collaboration_decision_2014}. They further facilitate evaluation of treatment options and promote healthier behavior and risk avoidance \cite{cox_behavioral_2010,fagerlin_making_2007,zikmund-fisher_communicating_2008}. Importantly, visual aids are consistently perceived as easier to understand, faster to process, and more memorable than text-based formats, again, particularly for vulnerable populations \cite{feldman-stewart_perception_2000,goodyear-smith_patients_2008,gaissmaier_numbers_2012}. Visual aids for non-expert users such as patients often include pictograms, icons or simple comics for orientation. These representations have been demonstrated to improve understanding and recall, and have a high acceptance among patients and clinicians across many disease entities \cite{delp_communicating_1996,dowse_pharmacists_2021,dowse_evaluation_2001,reeves_development_2023,shrestha_preparation_2018}. Despite these advantages, clinical communication remains largely text- and verbally driven; this is especially notable in oncology, where treatment pathways are complex, multistage, and associated with high uncertainty.
\section{Medical background}
\subsection{Cancer as a Multistage Treatment Process}
Cancer comprises a genetically heterogeneous group of diseases characterized by cell growth and the potential to metastases \cite{ferlay_global_2024,ottaiano_chaos_2023}. Reflecting the heterogeneity, treatment trajectories vary on biomarkers, stage, and patient condition, and usually involve a sequence of interventions, often including chemo-, radio-, surgical or transplant therapy. Interventions are organized into multiple phases, an induction phase aimed at reducing tumor burden, followed by consolidation or maintenance phases to prevent relapse and they usually follow trialled procedures that are harmonized across healthcare systems. From a communication perspective, cancer treatment can therefore be understood as a multistage temporal process, involving sequences of events, transitions between states, and periods of uncertainty. Patients must not only understand individual interventions but also how these are structured over time and how they relate to overall treatment goals \cite{epstein_patient-centered_2007}. This temporal and structural complexity makes cancer treatment plans particularly challenging to communicate verbally or with text alone.
\subsection{Leukemia}
Leukemia is a group of hematological malignancies affecting blood-forming tissues, including the bone marrow and lymphatic system \cite{miranda-filho_epidemiological_2018}. Acute myeloid leukemia (AML) is an aggressive subtype characterized by the rapid proliferation of abnormal myeloid cells \cite{heuser_acute_2020}. AML incidence increases with age, with a median age at diagnosis of ~70 years, and is expected to rise further due to population aging, as reflected by increasing incidence rates in Europe from 3.48 to 5.06 cases per 100,000 between 1976 and 2013 \cite{heuser_acute_2020}. AML is treatment usually is initiated with intensive outpatient induction chemotherapy, hospital-based high-dose chemomtherapy and transplantation with autologous stem cells, consolidation therapy, and long-term follow-up. Depending on patients biomarker composition, side-effects and therapy response, treatment pathways are adjusted. In addition, improved early detection and monitoring of disease recurrence have expanded the range of clinical presentations, further complicating leukemia treatment strategies \cite{shrestha_preparation_2018}. For patients, understanding the treatment trajectories and purpose of these phases is essential for informed decision-making but is often difficult due to the complexity of the information presented.
\subsection{Patient Demographics and Communication Challenges}
Patients undergoing cancer treatment represent a highly heterogeneous population with respect to age, education, language proficiency, and cognitive capacity. A large proportion of patients are older adults, many of whom experience age-related cognitive decline or treatment-related impairments that affect memory and comprehension \cite{jansen_recall_2008,kessels_patients_2003}. Additionally, clinical populations may include non-native speakers and individuals with limited health literacy, further complicating communication. Health literacy encompasses not only reading ability but also numeracy and the capacity to interpret complex, often probabilistic information \cite{lipkus_numeric_2007,fagerlin_making_2007}. In practice, many patients struggle to understand medical terminology, numerical risk information, and temporally structured treatment plans. These challenges are compounded by the emotional stress associated with a cancer diagnosis, which can further impair information processing and recall \cite{goerling_information_2020,husson_relation_2011}.
Despite the central role of communication in patient-centered care, substantial gaps remain in patients’ understanding and recall of information from their medical consultation \cite{ha_doctor-patient_2010}. In oncology, this problem is particularly severe, with some studies reporting recall rates as low as 25\% for treatment-related information \cite{faller_satisfaction_2016,godwin_they_2000,jenkins_what_2011}. At the same time, patients report a strong desire for clear, structured, and accessible information about their treatment pathways \cite{manta_patient_2021}. However, current communication practices, relying on verbal explanations by healthcare professionals and legal consent documents, are insufficient to address the complexity of multistage cancer treatments. Thus, this presents a clear unmet need for communication tools that translate complex treatment trajectories into formats that are both accurate and accessible. Visualization-based approaches, particularly those that explicitly represent temporal structure, offer a promising direction for addressing this gap.
\section{Design, UX optimization}
\subsection{Evaluation of baseline stimulus}
To improve comprehensibility and reduce cognitive load for non-expert users, the baseline AML treatment timeline (Figure~\ref{fig:1}a), which is based on harmonized treatments, was systematically evaluated and subsequently redesigned into two new versions: a visually optimized static representation and an interactive extension. Visualization and UX principles are particularly relevant in this context, as they are explicitly designed to support users without domain expertise by reducing complexity, structuring information, and aligning representations with human perceptual and cognitive processes. In contrast to expert-oriented systems, which assume prior knowledge and analytical motivation, patient-facing visualizations must prioritize clarity, guidance, and accessibility. The redesign was grounded in two frameworks, the Swiss Accessibility Checklist \cite{noauthor_accessibility_nodate} and the heuristics by Dowding and Merrill for healthcare visualization \cite{dowding_development_2018}. Applying the accessibility checklist with a designer-focused filter resulted in 49 relevant criteria, of which only 14 were applicable to the static baseline. Of these, only one criterion was fulfilled, while 13 were not or only insufficiently implemented (Table~\ref{tab:evaluation}). Identified weaknesses included missing alternative text, insufficient heading structure, limited color and contrast accessibility, and the absence of adaptive or zoomable presentation. The heuristic evaluation assesses usability items, of which 29 were applicable to the baseline, 15 of which were at least partially fulfilled and 14 not or not fully implemented (Table~\ref{tab:evaluation}). While principles such as Aesthetic and Minimalist Design (4/7 items/criteria met) and Recognition rather than Recall (2/4 items/criteria met) were partially addressed, key usability dimensions such as User Control and Freedom (0/5items/criteria met) and Match between System and the Real World (1/5items/criteria met) showed substantial deficits. Severe issues included missing system status visibility, ambiguous pictograms, weak visual grouping, inconsistent color encoding, and unclear temporal labels. These findings directly informed the subsequent redesign.
\subsection{Visual and Interactive re-design}
\begin{figure}[htbp]
    \centering
    \includegraphics[width=\linewidth]{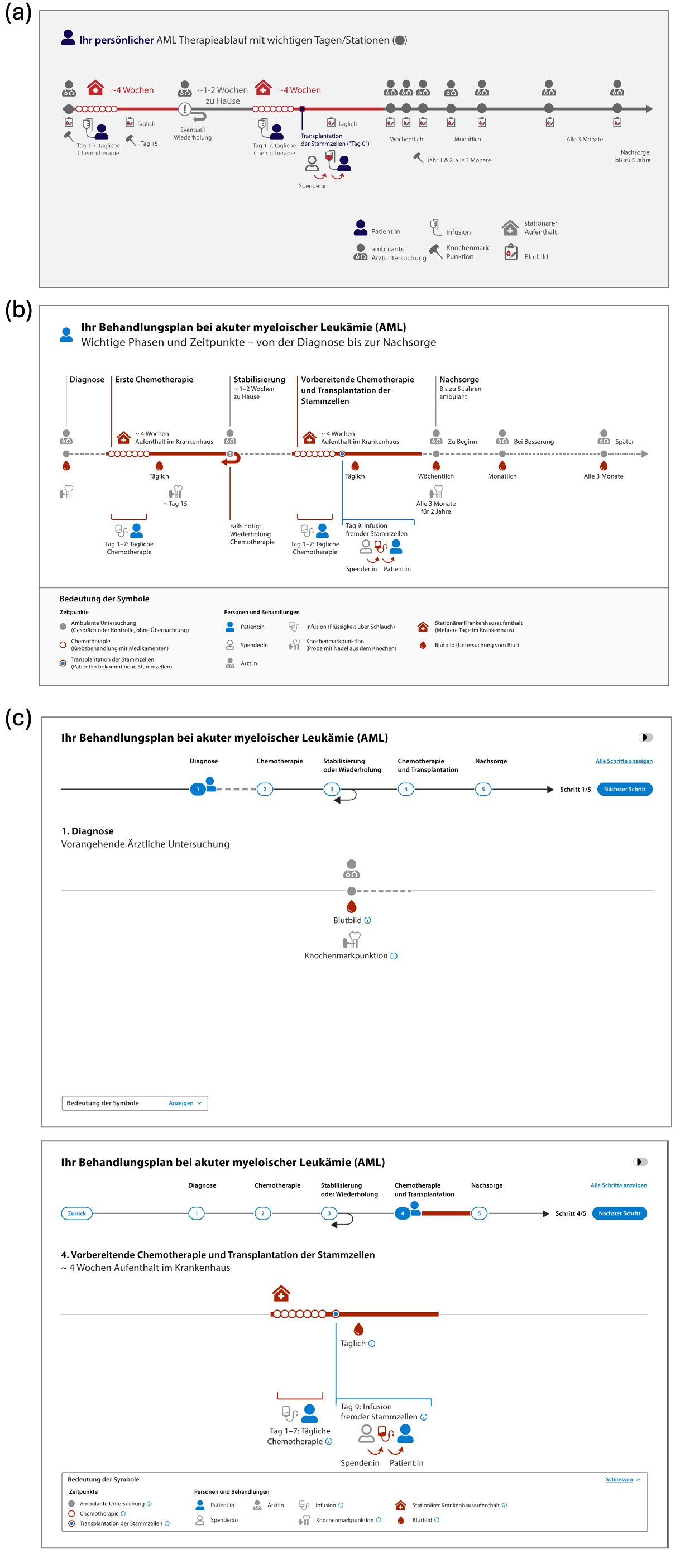}
    \caption{Overview of the AML treatment timeline designs. (a) Baseline AML timeline visualization showing a continuous sequence of treatment phases and follow-up intervals \cite{jambor_communicating_2025}. (b) Redesigned static timeline emphasizing key phases, durations, and clinical events with improved structure and labeling. (c) Interactive prototype presenting a step-wise navigation through treatment stages, allowing users to explore detailed information for individual phases. Icons represent clinical procedures, patient interactions, and temporal intervals.}
    \label{fig:1}
\end{figure}

The visual redesign (Figure~\ref{fig:1}b), focused on reduction, structure, consistency, and accessibility. Gestalt principles—proximity, similarity, and closure—were used to reorganize content into coherent units \cite{buhler_gestaltgesetze_2017}, supported by a consistent grid system to improve alignment and visual balance. To reduce cognitive load, information was structured into manageable chunks, and hierarchy was strengthened through improved positioning of headings, labels, and legends. Pictograms were selectively revised to improve recognizability and stylistic consistency. Highly abstract icons were simplified where users were unlikely to have a prior mental model, supported by open-source icon libraries such as Health Icons and the Noun Project. Color and contrast were redesigned according to WCAG 2.1 AA standards \cite{kirkpatrick_web_2018}, using tools such as Viz Palette and contrast checkers \cite{lu_viz_nodate,siefke_simulation_nodate}. Color was used sparingly and combined with additional visual encodings to avoid ambiguity, and typography and wording were standardized and simplified to improve readability and accessibility \cite{yablonski_laws_2020}.
The interactive extension (Figure~\ref{fig:1}c), aimed to support structured exploration, guidance and pacing without increasing complexity, while maintaining the horizontal timeline layout. A high-fidelity prototype was developed in Figma, introducing progressive disclosure to allow users to access additional information on demand. Interaction design emphasized consistency, with stable control placement and explicit feedback on system state, addressing the heuristic of visibility of system status \cite{dowding_development_2018}. Interaction patterns such as Completeness Meter and Steps Left were used to support orientation and progress tracking, while the number of simultaneously visible options was reduced in line with established principles \cite{yablonski_laws_2020}. An overview-plus-detail strategy enabled both linear navigation and selective exploration \cite{aigner_visualization_2023}. Interactive elements were implemented with defined states and a consistent visual language to ensure usability and accessibility.
\section{Evaluation}
\subsection{Study design}
To evaluate whether visual and interactive redesign can improve, or at least preserve, the comprehensibility of oncological treatment information, we conducted a quantitative online study using a between-subjects design with two independent one-factor comparisons. In the first comparison, the redesigned AML timeline was evaluated against the baseline timeline. In the second comparison, the interactive version was evaluated against the redesigned static version. These comparisons were analyzed separately to allow clearer attribution of effects and to reduce confounding. The study did not test the isolated effect of individual UX rules or accessibility criteria, but rather the effect of complete design solutions. We evaluated the designs along three dimensions: answer accuracy, answer confidence, and subjective rating of the information. These were assessed using ten multiple-choice comprehension questions, confidence ratings after each answer, and one overall rating of information comprehensibility. The survey was conducted in German and implemented in LimeSurvey. It included a short introduction to AML, instructions for participants, random assignment to one of the three stimulus conditions, ten multiple-choice questions with randomized answer order, sociodemographic and control-variable questions, and automatic recording of response time. 

The survey was anonymous, no personal data were collected, and participants provided informed consent before taking part. Contact information for the researchers was also provided. Participants were recruited through multiple channels, including personal networks, survey boards, patient-related networks, and social media, using a snowball strategy to increase respondent diversity. Inclusion criteria were an age of at least 18 years and access to a desktop device. An a priori power analysis was conducted in G*Power 3.1.9.6. Assuming a medium effect size and using a one-sided independent-samples t-test, the target sample size was 36 participants per condition.
\begin{figure}[htbp]
    \centering
    \includegraphics[width=0.8\linewidth]{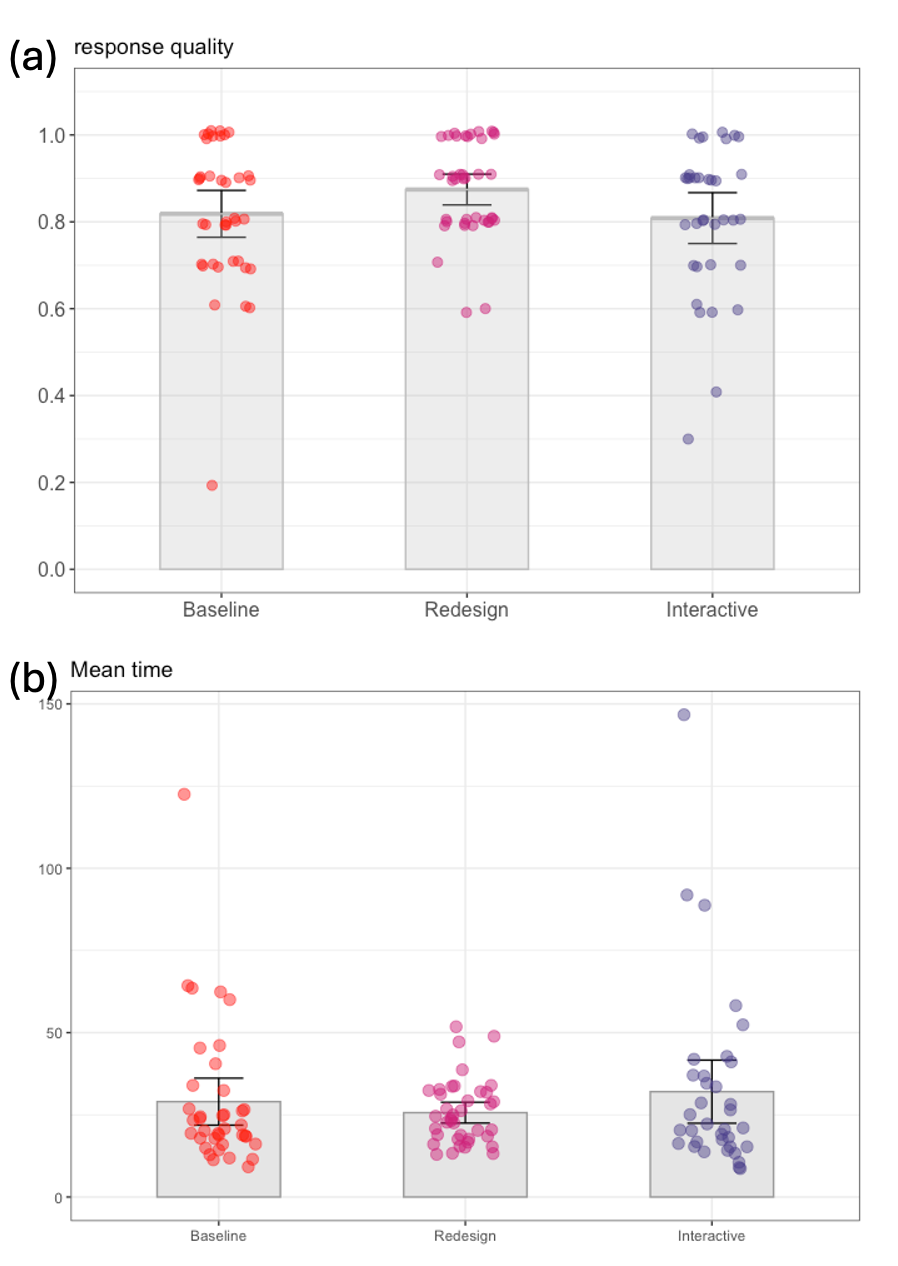}
    \caption{Participant performance for the baseline, redesign, and interactive conditions. Each point represents an individual participant, bars indicate group means, and error bars show variability across participants. (a) Mean response quality scores. (b) Mean response time (per question).}
    \label{fig:2}
\end{figure}
\subsection{Evaluation results}
A total of 112 participants were included in the analysis (Table~\ref{tab:socioeconomic}). Comprehension was assessed as the proportion of correct responses across the ten comprehension items (Fig.~\ref{fig:2}a). The redesigned timeline yielded the highest comprehension scores (M = 0.87, SD = 0.11), compared with the original timeline (M = 0.82, SD = 0.16) and the interactive prototype (M = 0.81, SD = 0.17). The redesigned timeline significantly improved comprehension relative to the original version, whereas the interactive prototype did not preserve this gain and performed worse than the redesigned static version. Thus, the redesign produced a measurable increase in participants ability to answer questions correctly about the AML treatment process, while the interactive extension did not. At item level, the redesigned static version showed consistently high success rates with comparatively low variability, with several items reaching 100\% correct responses. By comparison, the interactive condition showed somewhat lower success rates and greater variability. Exploratory covariance analyses further suggested that educational level was significantly associated with answer accuracy. In addition, mean response time (Fig.~\ref{fig:2}b) per comprehension item was shortest in the visual redesign condition (M = 25.65 s) and longest in the interactive condition (M = 32.04 s).

\begin{figure}[htbp]
    \centering
   \includegraphics[width=0.8\linewidth]{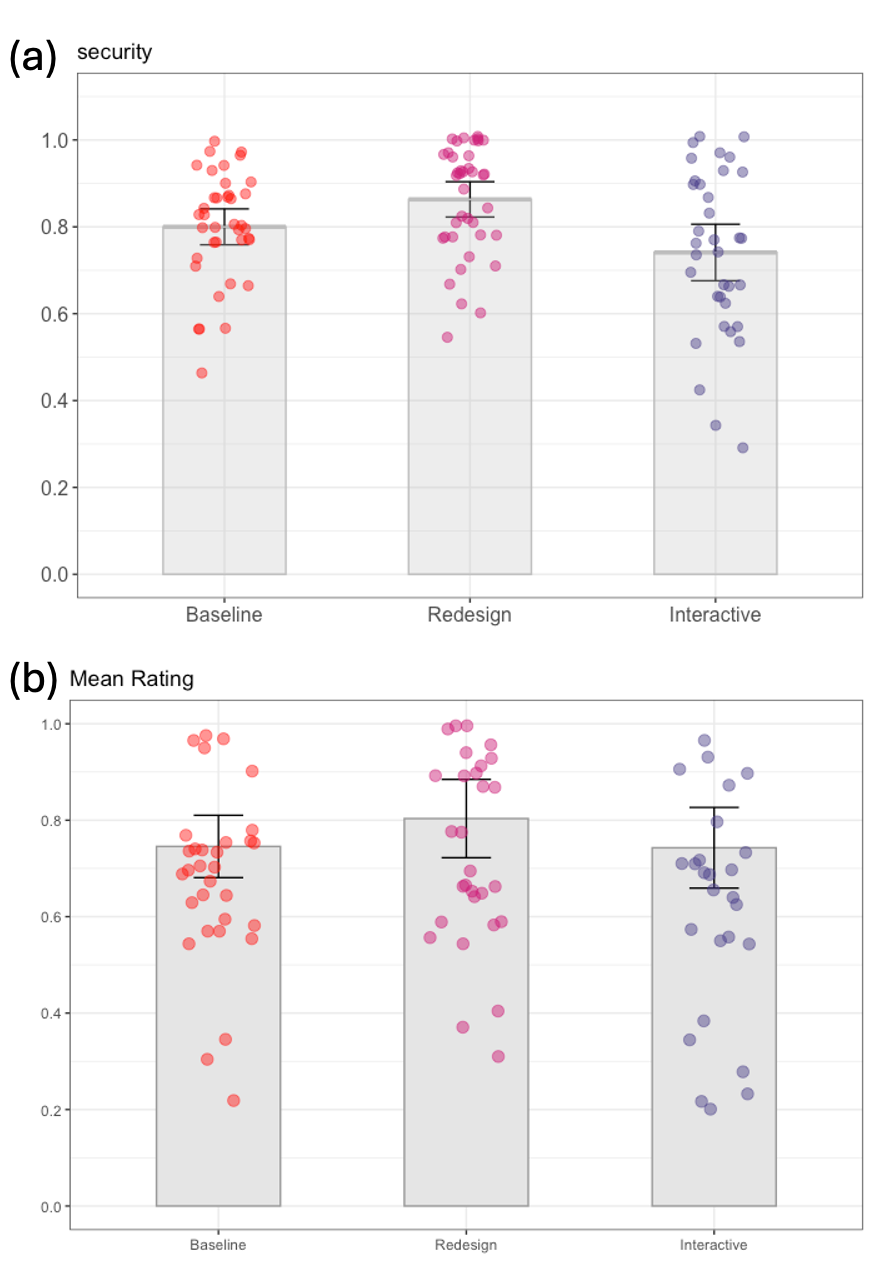}
    \caption{Subjective ratings of participants for baseline, redesign, and interactive conditions. Each point corresponds to an individual participant, with bars representing group means and error bars indicating variability. (a) Perceived confidence/security. (b) Overall mean ratings of information quality.}
    \label{fig:3}
\end{figure}

Beyond comprehension, participants rated their confidence after each comprehension item on a four-point scale, normalized from 0 - unsure, to 1 - sure (Fig.~\ref{fig:3}a). Again, the redesigned timeline produced the highest scores (M = 0.86, SD = 0.13), followed by the original condition (M = 0.80, SD = 0.13) and the interactive prototype (M = 0.74, SD = 0.19). Notably, the interactive prototype significantly reduced confidence relative to the redesigned static version, indicating inferiority of the interactive version with respect to confidence. This advantage of the redesigned timeline over the interactive prototype remained robust when controlling for age, prior knowledge, digital affinity, and visual impairment, suggesting that the observed difference was not driven by imbalance in the survey cohort. Participants were also asked to subjectively rate the overall comprehensibility of the provided information, including both the medical explanation and the treatment timeline (Fig.~\ref{fig:3}b). The visually redesigned condition again obtained the highest mean rating (M = 0.80, SD = 0.25), compared with the original timeline (M = 0.75, SD = 0.20) and the interactive prototype (M = 0.74, SD = 0.24). However, these differences did not reach statistical significance. Thus, while the visual redesign tended to receive the highest overall ratings, subjective evaluation of information quality remained broadly similar across conditions. Overall, both confidence and ratings, but not comprehension, were influenced by age, with older participants rating the information more positively than younger ones (61–75 years: M = 0.90, SD = 0.18; 18–30 years: M = 0.67, SD = 0.22).
\section{Discussion}

The results demonstrate that visual structure is a primary driver of understanding in the context of oncological treatment timelines. The redesigned timeline consistently improved comprehension and confidence compared to the original representation, indicating that relatively simple design interventions, such as clearer grouping, stronger visual hierarchy, and consistent encoding, can substantially improve interpretability. This aligns with prior work on health communication, which highlights that patients frequently struggle to understand complex medical information due to mismatches in literacy and numeracy \cite{lipkus_numeric_2007,galesic_using_2009,garcia-retamero_communicating_2013}. In such contexts, visual organization becomes critical, as it reduces the cognitive effort required to decode information. By structuring treatment phases into coherent units and aligning visual encodings with users’ expectations, the redesign supports perceptual processing rather than analytical reconstruction. Thus, stringently applying design heuristics and accessibility criteria can profoundly improve communication with patients as non-expert users and likely increase the low recall previously observed \cite{kessels_patients_2003,ley_memory_1979,jenkins_what_2011}.

In contrast to the improvements observed for visual structure, the addition of interaction did not yield further benefits compared to baseline and, in some cases, reduced comprehension and confidence. This suggests that interactivity does not inherently improve understanding and may introduce additional cognitive demands. In our design, interaction was primarily used to stage the information revealed based on user input. Even such low-level interaction may increase cognitive load, and might be particularly profound for the patient cohort \cite{kessels_patients_2003}. Our finding is thus consistent with prior emphasis placed on clarity, guidance, and simplicity over exploratory flexibility \cite{kosara_presentation-oriented_2016,chen_reflections_2020}. While narrative and interactive visualization approaches can enhance engagement, their effectiveness will depends strongly depend on how interaction is designed and integrated.

The findings have direct implications for the design of patient facing medical visualizations. In clinical practice, communication is still largely based on verbal explanations and text heavy materials, despite well documented limitations in patient comprehension and recall \cite{houts_role_2006,kessels_patients_2003}. This gap is particularly critical in oncology, where patients must understand multistage treatment pathways that involve uncertainty, risk, and long term planning. Our results suggest that structured visual timelines can effectively address these challenges by making treatment timelines explicit and reducing complexity. As healthcare systems face increasing resource constraints while treatments and stratification approaches become more personalized \cite{von_bonin_clonal_2021}, the importance of such tools is likely to grow. Participants also reported higher answer confidence and improved ratings of information quality. This is particularly relevant in light of a broader shift toward greater patient involvement through patient reported outcomes and experiences. Approaches that make temporal treatment trajectories more accessible, through visual aids or also narrative storytelling \cite{mittenentzwei_investigating_2023} therefore represent an important step toward more patient centered communication in healthcare, which remains a central goal of high quality care \cite{the_cochrane_collaboration_decision_2014,wolfe_institute_2001}.

This study has several limitations. First, the evaluation was conducted in an online setting, which may not fully reflect real-world clinical contexts. Participants differed from actual patients in terms of motivation, emotional involvement, and situational stress. However, prior work with similar visual aids consistently indicates that structured visual representations improve patient understanding, suggesting that the observed effects are likely transferable to clinical settings \cite{jambor_communicating_2025,dowse_evaluation_2001}. Second, the interaction design was intentionally constrained, more advanced forms of interaction, such as adaptive guidance, personalized pathways, or narrative-driven exploration, may yield different results and should be investigated in future work. Third, the study focused on a specific medical domain, namely a specific blood cancer, AML. While likely transferable to other multistage treatment processes, further studies are required to validate generalizability across different diseases, patient populations, and visualization types.
\section{Conclusions}
We presented and evaluated patient-centered visualizations of multistage cancer treatment trajectories, comparing a visually optimized static timeline with an interactive variant. The results show that improving visual structure with organization, hierarchy, and encoding based on established heuristics and criteria, leads to measurable gains in comprehension and confidence. In contrast, the addition of interaction did not improve understanding and, in this implementation, introduced additional cognitive demands. These findings emphasize that effective communication for non-expert users depends primarily on reducing interpretive effort rather than increasing functionality. In the context of patient communication, well-structured visual timelines can serve as powerful tools to convey complex temporal processes, support understanding, and facilitate informed decision-making. Future work should explore how interaction can be more effectively integrated, for example through guided or adaptive mechanisms, and how such visualizations perform in real clinical settings. More broadly, this work contributes to bridging the gap between visualization research and patient-centered healthcare, demonstrating how principles from information visualization and UX design can be translated into tangible improvements in medical communication.

\section*{Acknowledgements}
We thank the anonymous survey participants for their responses.
\section*{Data availability}
The interactive prototype of the AML treatment schedule can be accessed via the following link \url{https://www.figma.com/proto/F7QBDEUQWsGUh7r4RNAQ9V/AML-Behandlungsplan?page-id=222%3A1278&node-id=166-2165&viewport=-4352%2C104%2C0.22&t=JluxTIyj3W2WMSM5-8&scaling=contain&content-scaling=fixed&starting-point-node-id=166%3A2165&disable-default-keyboard-nav=1&hide-ui=1}  
\section*{Declaration of competing interests}
No competing interest.

\bibliographystyle{cag-num-names}
\bibliography{Laura-VCBM2}

\begin{table*}[t]
\caption{Usability evaluation.}
\label{tab:evaluation}
\begin{tabular}{l p{8cm} r}
\hline
\textbf{Usability factor} & \textbf{Item} & \textbf{Condition} \\
\hline
\multicolumn{3}{l}{\textbf{Swiss Accessibility Checklist}} \\
\hline
Lists & Lists with only one item are avoided (unless they are generated automatically). & 1 \\
Alt-text & If alternative text is not sufficient (e.g., for complex graphics), a long description is provided and referenced. & 0 \\
Headings & The hierarchy of heading levels is logically correct and conveys structure. No levels are skipped. & 0 \\
Headings & No heading levels are skipped. & 0 \\
Headings & Headings are followed by content or subordinate headings. & 0 \\
Layout & Content is usable in both screen orientations; if not, a manual rotation control is provided. & 0 \\
Color & Information is not conveyed by color alone; additional cues are provided. & 0 \\
Contrast & Text contrast is at least 4.5:1 (normal) or 3:1 (large text). & 0 \\
Text & Elements can be zoomed to at least 200\%. & 0 \\
Reflow & Content can be displayed without loss at 320×256 CSS pixels (400\% zoom). & 0 \\
Image contrast & Graphical elements have contrast ratio of at least 3:1. & 0 \\
Text sizing & Spacing can be increased without loss of functionality. & 0 \\
Headings & Headings and labels are descriptive and unique. & 0 \\
Consistent encoding & Components with the same function are implemented consistently. & 0 \\
\hline
\multicolumn{3}{l}{\textbf{Dowding \& Merrill Heuristic}} \\
\hline
Visibility of System Status & Does every screen have a descriptive title or header? & 1 \\
 & Is there a consistent icon design scheme? & 1 \\
 & Is there a clear indication of the current location? & 0 \\
 & Is menu terminology consistent with user tasks? & 1 \\
Match between System and Real World & Are icons concrete and familiar? & 0 \\
 & Are headings ordered logically? & 1 \\
 & Do colors correspond to expectations? & 0 \\
 & Are terms familiar to users? & 0 \\
Consistency and Standards & Are formatting standards applied consistently? & 0 \\
 & Are colors limited and well separated? & 0 \\
 & Are names consistent across the system? & 1 \\
 & Is color coding consistent? & 0 \\
Recognition rather than Recall & Are cues placed where users look? & 1 \\
 & Is whitespace used effectively? & 0 \\
 & Are items grouped logically? & 0 \\
 & Is highlighting used to guide attention? & 1 \\
Aesthetic and Minimalist Design & Is only essential information displayed? & 1 \\
 & Are layout elements used to distinguish sections? & 0 \\
 & Are labels brief and descriptive? & 0 \\
 & Is the layout well designed? & 1 \\
 & Are unnecessary elements avoided? & 1 \\
 & Is data presented simply? & 1 \\
 & Is whitespace used between color elements? & 0 \\
Spatial Organization & Are elements clear and visible? & 1 \\
 & Is the organization logical? & 1 \\
 & Is contextual detail provided? & 0 \\
Information Coding & Are symbols appropriate? & 1 \\
 & Are realistic characteristics used? & 1 \\
Orientation & Are measurement units clearly displayed? & 0 \\
\hline
\end{tabular}
\end{table*}

\begin{table*}[htbp]
\centering
\caption{Distribution of socio-economic characteristics and control variables. Numbers: responses, percent; M = mean, SD = standard deviation. Response time: seconds}
\label{tab:socioeconomic}
\begin{tabular}{lcccc}
 & Total & Baseline & Redesign & Interactive\\
\hline
\textbf{Age} \\
18--30 years & 32 & 14 (43.8\%) & 11 (34.4\%) & 7 (21.9\%) \\
31--45 years & 35 & 11 (31.4\%) & 8 (22.9\%) & 16 (45.7\%) \\
46--60 years & 16 & 4 (25.0\%) & 8 (50.0\%) & 4 (25.0\%) \\
61--75 years & 29 & 9 (31.0\%) & 12 (41.4\%) & 8 (27.6\%) \\
\hline
\textbf{Highest educational attainment} \\
No degree & -- & -- & -- & -- \\
Compulsory school & -- & -- & -- & -- \\
Vocational training & 20 & 10 (50.0\%) & 6 (30.0\%) & 4 (20.0\%) \\
High school diploma (Matura) & 14 & 6 (42.9\%) & 4 (28.6\%) & 4 (28.6\%) \\
Applied university & 34 & 8 (23.5\%) & 11 (32.4\%) & 15 (44.1\%) \\
University & 44 & 14 (31.8\%) & 18 (40.9\%) & 12 (27.3\%) \\
\hline
\textbf{Prior knowledge cancer} \\
No prior knowledge & 43 & 13 (30.2\%) & 11 (25.6\%) & 19 (44.2\%) \\
Some prior knowledge & 67 & 24 (35.8\%) & 27 (40.3\%) & 16 (23.9\%) \\
High prior knowledge & 2 & 1 (50.0\%) & 1 (50.0\%) & -- \\
\hline
\textbf{Digital affinity} \\
Not familiar & -- & -- & -- & -- \\
Rather unfamiliar & 6 & 2 (33.3\%) & 2 (33.3\%) & 2 (33.3\%) \\
Rather familiar & 40 & 16 (40.0\%) & 15 (37.5\%) & 9 (22.5\%) \\
Very familiar & 66 & 20 (30.3\%) & 22 (33.3\%) & 24 (36.4\%) \\
\hline
\textbf{Visual impairments} \\
Not sure & 1 & -- & 1 (100.0\%) & -- \\
No & 106 & 36 (34.0\%) & 38 (35.8\%) & 32 (30.2\%) \\
Yes & 5 & 2 (40.0\%) & -- & 3 (60.0\%) \\
\hline
\end{tabular}
\end{table*}

\end{document}